\documentclass[12pt,preprint]{aastex}

\def\f60fr{${\rm F_{60}}$/${\rm F_R}$}

\def\F60FB{${\rm F_{60}/F_{B}}$}
\def\F60F100{${\rm F_{60}/F_{100}}$}

\newcommand{\Ha}{\ensuremath{\mbox{H}\alpha}}%
\newcommand{\Hb}{\ensuremath{\mbox{H}\beta}}%
\def\L60{L$_{60}$}

\def\HII{\ion{H}{2}}
\def\ratioR23{([\ion{O}{2}]~$\lambda$3727 + 
[\ion{O}{3}]~$\lambda\lambda$4959,5007)/H$\beta$}
\def\R23{${\rm R}_{23}$}

\def\HI{\ion{H}{1}}

\shorttitle{}
\shortauthors{}

\begin{document}

\title{The H$\alpha$ and Infrared Star Formation Rates
for the Nearby Field Galaxy Survey}

\author{Lisa J. Kewley\altaffilmark{1}}
\affil {Harvard-Smithsonian Center for Astrophysics}
\authoraddr{ 60 Garden Street MS-20, Cambridge, MA 02138}
\email {lkewley@cfa.harvard.edu}
\altaffiltext{1}{CfA Fellow}

\author{Margaret J. Geller}
\affil{Smithsonian Astrophysical Observatory}
\authoraddr{ 60 Garden Street MS-20, Cambridge, MA 02138}

\author{Rolf A. Jansen}
\affil {Dept. of Physics \& Astronomy, Arizona State University}
\authoraddr{P.O. Box 871504, Tempe, AZ 85287-1504}

\author{Michael A. Dopita}
\affil {Research School of Astronomy and Astrophysics, Australian National
	University}

\begin{abstract}
We investigate the \Ha\ and infrared star formation rate (SFR)
diagnostics for galaxies in the Nearby Field Galaxy Survey (NFGS). 
For the 81 galaxies in our sample, we derive \Ha\ fluxes 
(included here) from integrated spectra.  
There is a strong correlation between the ratio of far-infrared to optical
luminosities L(FIR)/L(\Ha) and the extinction E($B-V$)
measured with the Balmer decrement.  
Before reddening correction, the SFR(IR) and 
SFR(\Ha) are related to each other by a power-law:
${\rm SFR(IR)} = (2.7 \pm 0.3)\, {\rm SFR(H\alpha)}^{1.30 \pm 0.06}$.
Correction of the SFR(\Ha) for
extinction using the Balmer decrement and a classical reddening curve
both reduces the scatter in the SFR(IR)-SFR(\Ha) correlation and results
in a much closer agreement between the two SFR indicators;
${\rm SFR(IR)} = (0.91 \pm 0.04) \,
{\rm SFR (H \alpha_{corr})^{1.07\pm0.03}}$.  

SFR(IR) and SFR(\Ha)
agree to $\sim 10$\%.
This SFR relationship spans 4 orders of magnitude and holds for all 
Hubble types with IRAS detections in the NFGS.    A constant ratio 
between the SFR(IR) and
SFR(\Ha) for all Hubble types, including early types (S0-Sab), 
suggests that the IR emission in all of these objects
results from a young stellar population. 
\end{abstract}


\section{Introduction}

Understanding the star formation history of the universe is the
primary goal of much current research in astronomy  \citep[for
example,][]{Rowan-Robinson01,Cole01,Baldry02,Lanzetta02,
Rosa-Gonzalez02}.
To obtain this understanding, a reliable estimate of the star
formation rate in individual galaxies is required.  Many calibrations
of star-formation rate depend on the luminosity measured
at various wavelengths including: radio
\citep[eg.,][]{Condon92,Cram98,Haarsma00},  IR
\citep[eg.,][]{Hunter86,Lehnert96,Meurer97,Kennicutt98},  optical
\citep[eg.,][]{Gallagher89,Leitherer95a,Kennicutt98,Madau98,
Rosa-Gonzalez02}, or  UV
\citep[eg.,][]{Buat89,Deharveng94,Leitherer95b,Meurer95,Cowie97,Madau98,
Rosa-Gonzalez02}.  Unfortunately, the agreement among the SFR
indicators at these different wavelengths is poor, and the underlying reasons
for the differences are not well understood.  To compute the SFR
at any redshift and to relate SFRs at different redshifts, it is
crucial to understand the physics underlying each SFR indicator and
the source of the discrepancies among them.

Here, we investigate two SFR indicators, the SFR(IR) and SFR(\Ha), for
a large, objectively selected sample of nearby galaxies.  We aim
to: (1) reduce the discrepancy between these two SFRs, and (2) obtain
a better understanding of 
the relationship between the IR and \Ha\ emission for all Hubble types.

In star-forming galaxies, \Ha\ photons are produced by gas ionized
by young hot stars.  The SFR(\Ha) is then a direct measure of the current 
SFR in galaxies, provided that reddening is not significant. 
The SFR(\Ha) should agree with the SFR(IR) if the
IR emission in star-forming galaxies arises from
dust heated by the young hot stars.  This scenario occurs in IR luminous
starbursts \citep[eg.][]{Dopita02}, but the IR emission 
in normal star-forming galaxies appears more complex.

The IR emission (1--1000$\mu$m) in normal star-forming galaxies may arise 
from three processes: (1) the emission from dust heated by young OB stars
\citep{Devereux90,Devereux97}; (2) the emission from the photospheres or
circumstellar envelopes of evolved stars such as red-giants undergoing
mass-loss \citep{Knapp92,Mazzei94}; (3) the ``cirrus'' emission from
dust distributed throughout the optically thin, neutral interstellar
medium, heated by the general stellar radiation field.  The general
stellar radiation field is composed of a combination of unabsorbed
radiation from young OB stars inside the active star-forming regions
\citep[eg.,][]{Zurita00}, and an underlying old stellar population
consisting of A, F and G dwarfs, and K and M giants
\citep[eg.,][]{Drapatz79, Helou86,Lonsdale87,Sauvage94,Mazzei94}.  The
relative contribution to the general stellar radiation field by the
young and old stellar populations is unclear and may vary substantially
from galaxy to galaxy depending on the details of the physical and
dynamical environment and metallicity. 

The calibration of infrared luminosity as an indicator of the global
star formation rate in a galaxy relies upon two assumptions; (a) that
young stars dominate the radiation field throughout the UV to the
visible (ie., that processes (2) and (3) above are negligible), and (b)
that the dust opacity is infinite throughout a galaxy.  If these
assumptions hold, then the IR luminosity approximates the bolometric
luminosity of the galaxy, and the SFR(IR) is a reliable estimate of the
true SFR of the galaxy. 

For infrared-bright dusty star-forming galaxies, the young stellar
population does produce the dominant dust heating radiation field, and
the dust opacity is high
\citep{Lonsdale87,Poggianti01,Dopita02,Rosa-Gonzalez02}. Therefore,
assumptions (a) and (b) above hold, and IR luminosity is 
a more direct indicator of the true SFR than the \Ha\ luminosity.
Recently, \citet{Dopita02} (hereafter D02) and \citet{Rosa-Gonzalez02} 
(herafter R02) showed that using the Balmer decrement to correct the 
\Ha\ luminosity for reddening results in better agreement between the 
SFR(IR) and SFR(\Ha).   The D02 sample contains merging infrared starburst
galaxies; the R02 sample consists of  nearby young star-forming
galaxies, most are low metallicity  late-type spirals or blue
compact dwarfs.  It is unclear from these results whether the approximate
agreement between SFR(IR) and SFR(\Ha) after reddening correction also
applies to a wider range of ``normal'' galaxies.

The young (OB) star-forming population dominates the IR emission in 
late-type spirals.  The nature of the IR emission
in  early-type spirals, on the other hand, remains 
controversial.   Dust heating by the general stellar radiation field may
make a significant  contribution to the IR emission in early-types
\citep[eg.,][]{Lonsdale87,Sauvage94,Mazzei94} bringing the use of
SFR(IR) into question for these galaxies.   Early-type galaxies may
have small dust opacities further complicating the use of SFR(IR).

\citet{Kennicutt83} demonstrated that the \Ha\ equivalent width
increases systematically from early-type to late-type galaxies.  This
increase suggests that early-type galaxies are deficient in young stars. 
\citet{Sauvage94} provide additional evidence for this deficiency.  They
found that the $B-V$ and $U-B$ colors decrease and the ratio of \HI\
mass to blue luminosity increases systematically from S0 to Im type in
the CfA1 galaxy sample \citep{Davis83,Huchra83}.  Furthermore, IR SED
modeling of early-types appears to require a cooler component to the IR
emission \citep{Buat88,Rowan-Robinson89}, often attributed to the
general stellar radiation field. 

An intriguing theoretical argument by \citet{Inoue02} suggests that
the SFR(IR) can be applied to a wide range of galaxy types, including
those with an older stellar population and small dust opacity.   Inoue
proposes a scenario in which the two effects of small dust opacity and
a large cirrus contribution offset each other, making the SFR(IR) 
a good measure of the true SFR within a factor of 2 for a wide range 
of galaxy types.

Although the general stellar radiation field may be important in many
early-type spirals, \Ha\ images provide convincing evidence that at
least some early-type galaxies contain significant recent star formation
\citep{Hameed99}.  In addition, \citet{Tomita96} and
\citet{Devereux97} showed that the far infrared-to-blue luminosity ratio
is independent of Hubble type for a large sample of nearby spirals.
Their work supports the idea that early-type galaxies contain substantial 
recent star formation. 

A number of studies compare various SFR indicators for normal disk
galaxies \citep[eg.,][]{Kennicutt83o,Hopkins01,Charlot02,Buat02}.  Most
of these authors cannot properly correct the \Ha\ luminosity for
reddening or stellar absorption because they lack integrated optical
spectra with
sufficient spatial coverage, signal-to-noise, and wavelength coverage. For example,
\citet{Charlot02} use  had to assume an `average' attenuation ${\rm A_{v}}=1$
for the galaxies in the Stromlo-APM survey where the slit-covering fraction
is 40-50 percent. 

In this paper, we analyze integrated spectra from the Nearby Field Galaxy Survey (NFGS)
\citep{Jansen00a}, described in Section~\ref{sample}.  We
 ensure that the integrated \Ha\ luminosity is properly corrected for
reddening and underlying stellar absorption.  We compare uncorrected
SFR(\Ha) with SFR(IR) in Section~\ref{SFR-uncorr}.  Section~\ref{SFR-corr}
compares the corrected SFR(\Ha) with the SFR(IR).  The
agreement between the two SFR indicators in the latter section is
remarkable and is independent of Hubble type.  We discuss the
implications of this agreement in Section~\ref{conclusions}.

\section{Sample selection and cross-correlation with IRAS \label{sample}}

The NFGS is ideal for investigating
relative IR and \Ha\ star formation rates.  The sample was selected
objectively from the CfA1 galaxy survey \citep{Davis83,Huchra83} with
${\rm m_{B(0)}} < 14.5$.  The CfA1 catalog is nearly complete within its
selection limits and contains galaxies with a large range of absolute
magnitude ($-22 \lesssim {\rm M_{z}} \lesssim -13$).  \citet{Jansen00a}
chose a subsample of 198 nearby galaxies spanning the full range of
Hubble type and absolute magnitude present in the CfA1 catalog.  To
avoid a strict diameter limit, which might introduce a bias against the
inclusion of low surface brightness galaxies in the sample, they chose a
radial velocity limit, ${\rm V_{LG} (km\,s^{-1}) > 10^{-0.19-0.2M_{z}}}$
(with respect to the Local Group standard of rest).  To avoid a sampling
bias favoring a cluster population, they excluded galaxies in the
direction of the Virgo Cluster. 
The absolute magnitude distribution in the NFGS sample approximates the
local galaxy luminosity function, while the distribution over Hubble
type follows the changing mix of morphological types as a function of
luminosity in the local galaxy population. Low surface brightness galaxies
are overrepresented in the NFGS relative to the CfA1 sample.

Both integrated and nuclear spectrophotometry are available for almost all
galaxies in the NFGS sample, including reliably measured integrated \Ha\
and \Hb\ fluxes relative to the flux at 5500\AA\ \citep{Jansen00b,Jansen01}.
The spectra have high enough S/N ratios to measure the \Hb\ fluxes
even when E(B-V) is large.

The integrated spectra typically cover 82$\pm$ 7\% of the galaxy.
We calibrate the integrated fluxes, ${\rm F_{rel}}(\lambda)$,
to absolute 
fluxes, ${\rm F}(\lambda)$, by careful comparison with B-band
photometry.
We converted total $B$-magnitudes (extrapolated to infinite radius) to
flux densities, ${\rm F(4350\AA,phot)}$, at the effective wavelength of
the $B$-filter (4350\AA).  The uncertainties in the constant relating
the magnitude to the flux of Vega, and in the effective wavelength of
the $B$-filter for a given galaxy SED are $\sim$2-3\% combined.  Because the
position, size and orientation of the spectroscopic apertures are known,
the $B$-filter galaxy images allow us to compute the fractions, $f,$  
of the total $B$-filter light that enters the spectroscopic apertures with
1.5-3\% accuracy.  The mean value of $f$ = $0.82\pm0.07$. The
$B$-filter flux density to which we must scale
the spectra is then
\begin{displaymath}
  {\rm F(}B,{\rm spec)} = f \times {\rm F(4350\AA,phot)}.
\end{displaymath}
Scaling the relatively flux calibrated spectrum, ${\rm
F_{rel}}(\lambda)$, by the ratio of this flux density and that derived
from the spectrum itself, we obtain the absolute flux calibrated
spectrum:
\begin{displaymath}
   {\rm F}(\lambda) = {\rm F_{rel}}(\lambda) \cdot
      \frac{{\rm F(}B,{\rm spec)}}{\langle{\rm F_{rel}}(\lambda) \otimes
      {\rm T}_B(\lambda)\rangle}\quad\
      {\rm [erg\,s^{-1} cm^{-2} \AA^{-1}]}\ ,
\end{displaymath}
where $\langle{\rm F_{rel}}(\lambda) \otimes {\rm T}_B(\lambda)\rangle$
denotes the mean relative flux density, derived by convolving the
relative spectrum with the normalized $B$-filter response curve (Bessell
1979).  The maximum error in the relative flux calibration is
$\sim$6.5\%.  The range in uncertainty in the total $B$-filter
magnitudes is $\sim$1-10\%, with a mean of 3.0\% and rms of 2.7\%.  The
errors in the absolute flux calibration of the spectrum are smallest at
4350\AA\ and increase toward both shorter and longer wavelengths.  The
expected error in the derived absolute \Ha\ (\Hb) emission line fluxes
is on average $\sim$12\% ($\sim$17\%) and at most $\sim$34\%
($\sim$50\%) for galaxies with ${\rm EW(\Hb)}\lesssim -0.5${\AA}.  The
error in the \Hb\ flux is larger than that in \Ha, because the \Hb\
emission line is fainter and its measurement is complicated by the
superposition of stellar absorpion. 

As described in \citet{Jansen01}, the line fluxes are corrected for
Galactic extinction, using Burnstein \& Heiles (1984; as listed in
de~Vaucouleurs et al.\ 1991).  The \Hb\ emission is partially corrected
for stellar absorption by ensuring that the limits of the flux
measurement window are well inside the absorption trough, closely
bracketing the emission line.  We evaluated the residual absorption
using spectra of galaxies without detectable emission.  On average, we
applied an additional correction of 1.0\AA\ in equivalent width (EW) to
\Hb\ and 1.5\AA\ EW to \Ha. 

We cross-correlated the NFGS with the IRAS Faint Source Catalogue (FSC)
and Point Source Catalogue (PSC).  The IRAS beam size at 60~$\mu$m is
1.5 arcminutes, and the IRAS positional uncertainty is 30 arcseconds. 
Because the NFGS positional uncertainty is only a few arcseconds and
because all but a few of the NFGS galaxies are smaller than 3~arcmin in
extent, we used a detection radius of 30 arcseconds.  Any IRAS source
associated with an NFGS object should be detected within this radius.  We
found 110 (56\%) NFGS galaxies with IRAS sources 
within 30 arcseconds; 84 of these galaxies have moderate or good quality 
fluxes at 60 and 100$\mu$m, and \Ha\ and \Hb\ fluxes accurate to 
within 30\%.

To rule out the presence of AGN in the NFGS sample, we used the
theoretical optical classification scheme developed by \citet{Kewley01}. 
The optical diagnostic diagrams indicate that the global spectra of
81/84 NFGS galaxies are dominated by star formation.  The spectra of 
the remaining 3 galaxies are dominated by AGN: UGC 00545, 
Mrk 0205, and NGC~5940 host Seyfert~{\sc i} nuclei.
These 81 galaxies 
constitute the sample we analyse here.  We list the galaxies in
Table~\ref{sample_table}.

\section{Infrared versus \boldmath\Ha\ SFRs \label{SFR-uncorr}}

The development of SFR diagnostics has been an intense topic of
research for over three decades \citep[see][for a review]{Kennicutt98}
(hereafter K98). We use the K98 SFR relations for IR and \Ha\ because 
these relations are used in many current SFR studies.  We defer a more
detailed  analysis of other SFR diagnostics to a future paper.

The IR SFR relation (K98) depends on the total infrared luminosity 
(8-1000$\mu$m) (${\rm L_{IR}}$):

\begin{equation}
{\rm SFR(IR) (M_{\odot} yr^{-1}) = 4.5\times10^{-44}\, L_{IR}\, (ergs\,s^{-1})}.
\end{equation}
Note that ${\rm L_{IR}}$ 
is referred to as ${\rm L_{FIR}}$ in K98.  For the remainder of this paper, 
we refer to the 40-120$\mu$m range as the far-infrared (FIR), and 
the 1-1000$\mu$m range as the infrared (IR) regime.

The K98 SFR relation assumes that young stars dominate the radiation 
field throughout the UV-visible
regimes, and that the optical depth is large. The infrared luminosity
is then a good representation of the bolometric luminosity of the
galaxy.  Although this situation may hold for IR-luminous starburst
galaxies, it does not necessarily apply to normal galaxies where the
emission may be dominated by emission from a cooler,  ``infrared
cirrus'' component \citep[eg.,][]{Lonsdale87}.

In order to estimate the infrared star formation rate SFR(IR), we first 
calculate the far-infrared 
flux (F(FIR): 40-120$\mu$m) from the IRAS 60 and 100$\mu$m fluxes 
(${\rm F}_{60}, {\rm F}_{100}$) in units of
Janskies \citep{Helou88}:

\begin{equation}
{\rm F(FIR)}=1.26\times 10^{-14}[2.58 {\rm F}_{60} + {\rm F}_{100}]\,\,
({\rm W\,m^{-2}})
\end{equation}  
 
\noindent
The far-infrared luminosity L(FIR) is then calculated using:

\begin{equation}
{\rm L(FIR)= 4 \pi D^{2}_{L} F(FIR) }
\end{equation} 
where D$_{\rm L}$ is the cosmological ''luminosity distance".  
We assume H$_{0}$=75 km s$^{-1}$ Mpc$^{-1}$ 
and q$_{0}$ = 0.5. For thermal energy distributions relevant for most galaxies, the FIR flux 
and luminosity are essentially model independent.  All galaxies in our
sample have measured 60 and 100$\mu$m fluxes.
 
New IR photometry and spectra from the
Infrared Space Observatory (ISO) have lead to recent advances in 
infrared spectral energy distribution (SED) modelling
\citep[eg.,][]{Spinoglio02,Dale01,Rowan-Robinson01,Calzetti00,Granato00}.
It is now possible to
obtain estimates of the total infrared 
luminosity from the IRAS 60 and 100$\mu$m fluxes 
\citep[eg.,][]{Charlot02,Rosa-Gonzalez02}. 
The uncertainties that arise from the conversion of 60 and 100$\mu$m IRAS fluxes 
to total (1-1000$\mu$m) IR flux are likely to be modest in comparison with the 
uncertainties in the conversion from infrared
luminosity to star formation rate \citep[][]{Charlot02}.

To transform L(FIR) to L(IR), we use the \citet{Calzetti00} 
(hereafter C00) formulation because it was developed for IR-luminous 
starburst galaxies and is therefore the most
consistent relation to use in conjunction with the K98 SFR
formula.  C00 modelled the total-infrared flux for
five low-redshift starburst galaxies with a two-component dust model.
The dust model consists of two modified Planck functions of 
temperature T$\sim40-55$~K (warm dust) and T$\sim20-23$~K (cool dust),
with an emissivity index $\epsilon = 2$.  C00 finds that the ratio of
the total dust IR emission (1-1000$\mu$m) to the IRAS FIR emission is

\begin{equation}
{\rm L(IR)} \sim 1.75 \times {\rm L(FIR)}
\end{equation}

\noindent
with little variation from galaxy to galaxy.
Using this conversion, the K98 SFR(IR) becomes:

\begin{eqnarray}
{\rm SFR(IR)}\,({\rm M_{\odot} yr^{-1}}) & = & 4.5\times10^{-44}\, {\rm L(IR)} \nonumber \\
                                        & \approx & 7.9 \times10^{-44}\, {\rm L(FIR)}\, ({\rm ergs\,s^{-1}}) 
\end{eqnarray}

Here, we define SFR(IR) as the SFR derived from the
total (1-1000$\mu$m) IR luminosity.   We estimate the IR luminosity from the measured
IRAS 60 and 100$\mu$m fluxes.  We note that K98 defined the SFR(IR)
over 8-1000$\mu$m.  The C00 conversion applies to the range 1-1000$\mu$m.
The contribution to the total infrared luminosity from the 1-8$\mu$m
regime is expected to be of the order of a few percent
\citep{Calzetti00,Dale01}. 

The \Ha\ SFR relation (K98) comes from evolutionary
synthesis models, assuming solar metallicity and no dust.  The  total
integrated stellar luminosity shortward of  the Lyman limit is
re-emitted in the nebular emission lines. The \Ha\ SFR relation is:
\begin{equation}
{\rm SFR(H\alpha) (M_{\odot} yr^{-1}) = 7.9 \times 10^{-42}\, L(H\alpha)\, (ergs\,s^{-1})}
\end{equation}
For most star-forming galaxies, dust
absorption leads to the underestimation of the SFR derived from the
\Ha\ luminosity  \citep[eg.,][]{Lonsdale87}.

\begin{figure}
\plotone{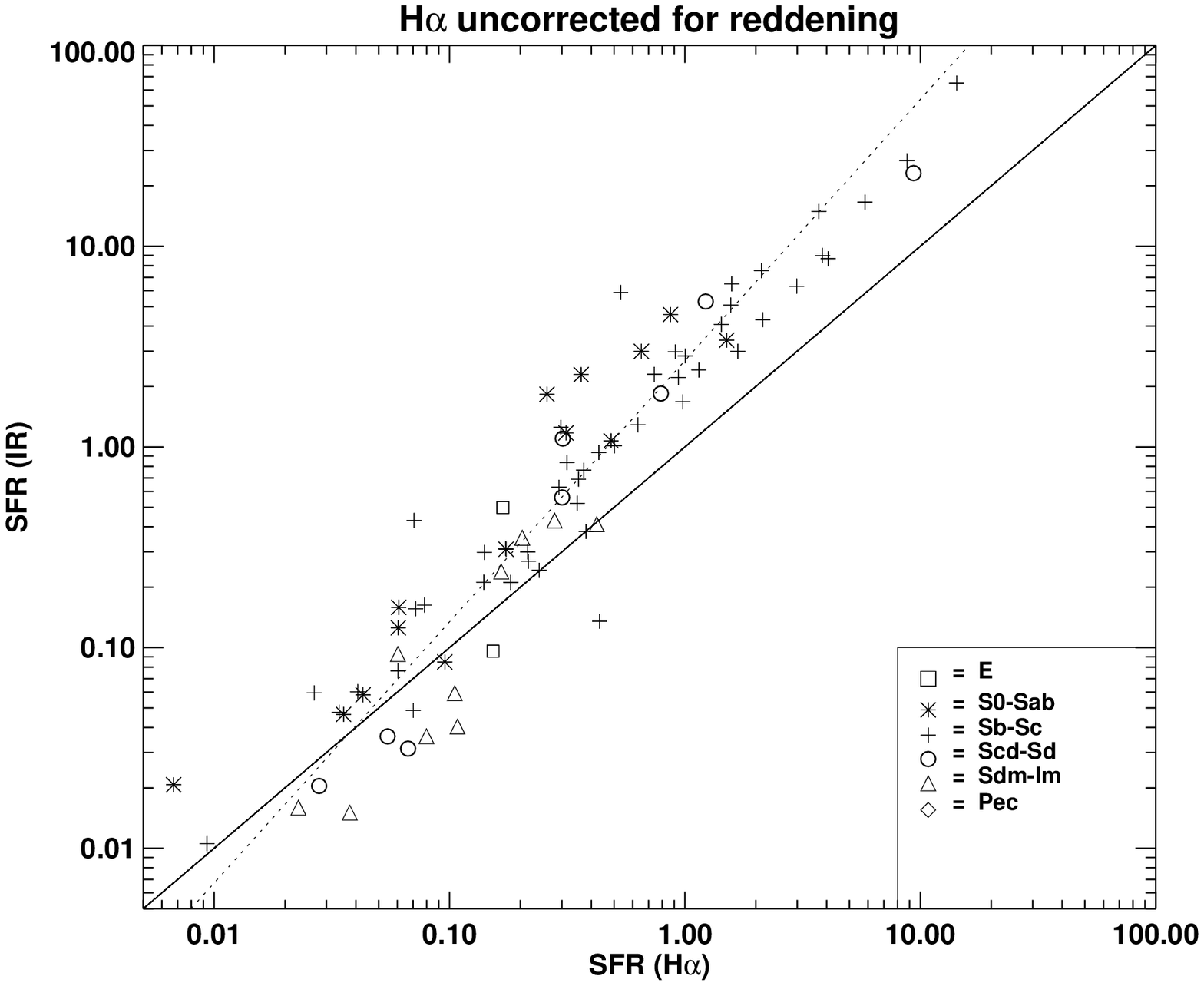}
\caption{A comparison of the SFRs derived using the infrared and \Ha\ 
relations in \citet{Kennicutt98}.  The strong correlation 
spans 4 orders of magnitude.  The solid line is y=x, and shows
where the data would lie if both SFR indicators agreed.
The \Ha\ SFR has not been corrected for extinction.
The dashed line is the best-fit to all of the data. The legend
indicates the Hubble type.
\label{SFR_Ha_vs_SFR_IR}}
\end{figure}

Figure~\ref{SFR_Ha_vs_SFR_IR} shows the SFRs derived using the  K98 IR
and \Ha\ SFR relations for the NFGS galaxies.  There is obviously a 
strong correlation between the two SFRs.   Upper limits for those objects 
with measurable \Ha\ fluxes but without IRAS detections at 60 and 100$\mu$m
are consistent with this correlation.

\placefigure{SFR_Ha_vs_SFR_IR}
We fit a straight line to the logarithm of the SFRs 
for the objects with IRAS detections using the numerical recipes
\emph{fitexy} routine in IDL.  This routine uses linear least-squares
minimization and includes error estimates for both variables.  We assumed
errors of $\sim$60\% for the non-reddening corrected SFR(\Ha) and $\sim$30\%
for SFR(IR).  The large $\sim$60\% error allows for the systematic error
introduced by failure to correct for reddening.
The resulting fit (dotted line in Figure~\ref{SFR_Ha_vs_SFR_IR}) 
has the form:

\begin{equation}
\log[{\rm SFR(IR)}] = (1.30\pm0.06) \log[{\rm SFR(H\alpha)}] + 0.43\pm0.05
\end{equation}
or:
\begin{equation}
{\rm SFR(IR)} = (2.7 \pm 0.3){\rm SFR(H\alpha)}^{1.30 \pm 0.06}.\label{SFRpower_law}
\end{equation}
The rms distance perpendicular to this line (in the log) 
gives a measure of the degree of scatter.  
The rms dispersion in Figure~\ref{SFR_Ha_vs_SFR_IR} is 0.17 in the log.
The Spearman Rank correlation test for SFR(IR) and 
SFR(\Ha) gives a
correlation coefficient of 0.95. The two-sided probability of finding
a value of 0.95 by chance is formally 0.0 ($\lesssim 10^{-30}$), 
confirming the very strong correlation between the two SFR indicators.

\begin{figure}
\plotone{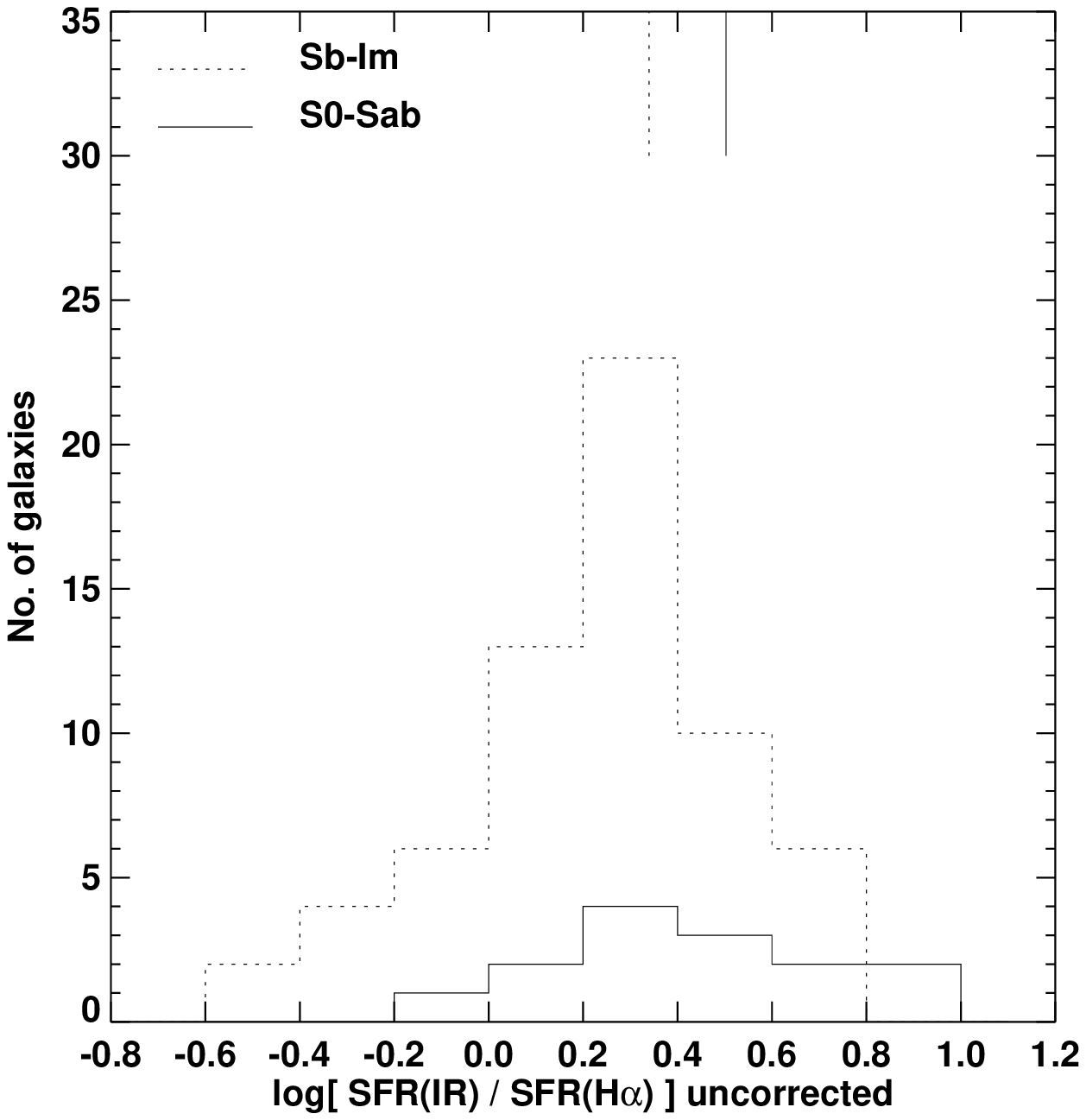}
\caption{The ratio of the infrared to \Ha\ star formation rates. The
\Ha\ flux has not been corrected for
reddening.  The two vertical lines at the top of the Figure indicate
the mean of the early and late-type distributions.
\label{hist_uncorr}}
\end{figure}

We plot a histogram of the logarithm of the ratio of the two SFRs, $\log[{\rm SFR(IR)/SFR(H\alpha)}]$,  
in Figure~\ref{hist_uncorr}.  Markers at
the top of Figure~\ref{hist_uncorr}  show the
means of the early and late
type SFR ratio distributions, $0.50\pm0.07$ and $0.34\pm0.04$, respectively.
We note that the mean for the early types is larger than the mean for
the late types. 

\placefigure{hist_uncorr}

\section{SFRs and Reddening \label{SFR-corr}}

Figures~\ref{SFR_Ha_vs_SFR_IR}~and~\ref{hist_uncorr}, and equation~\ref{SFRpower_law}
demonstrate that the SFR(IR) estimate exceeds SFR(\Ha) by 
a factor of $\sim 3$, as observed by \citet{Charlot02}.  The ratio 
between SFR(IR) and SFR(\Ha) increases with SFR.    
 If the SFR(IR) and SFR(\Ha) are both valid measures of the true
SFRs, then a physical process not yet taken into account must  bias
one relative to the other.  This process must also cause the bias to 
increase at higher SFRs.  One obvious candidate is reddening.  We
test this conjecture by  measuring the difference in the extinction at
two wavebands, E($B-V$).

We calculated E($B-V$) in the emission-line gas for the 81
IRAS-detected galaxies in the NFGS sample for which flux measurements
for both \Ha\ and \Hb\ are available.  We assume an intrinsic
\Ha$/$\Hb\ ratio of 2.85 (case B recombination at T$=10^{4}$K and
electron density $n_{e} \sim 10^{2} - 10^{4} {\rm cm}^{-3}$;
\citet{Osterbrock89}) and used the Whitford reddening curve as
parameterized by \citet{Miller72}.

\begin{figure}
\plotone{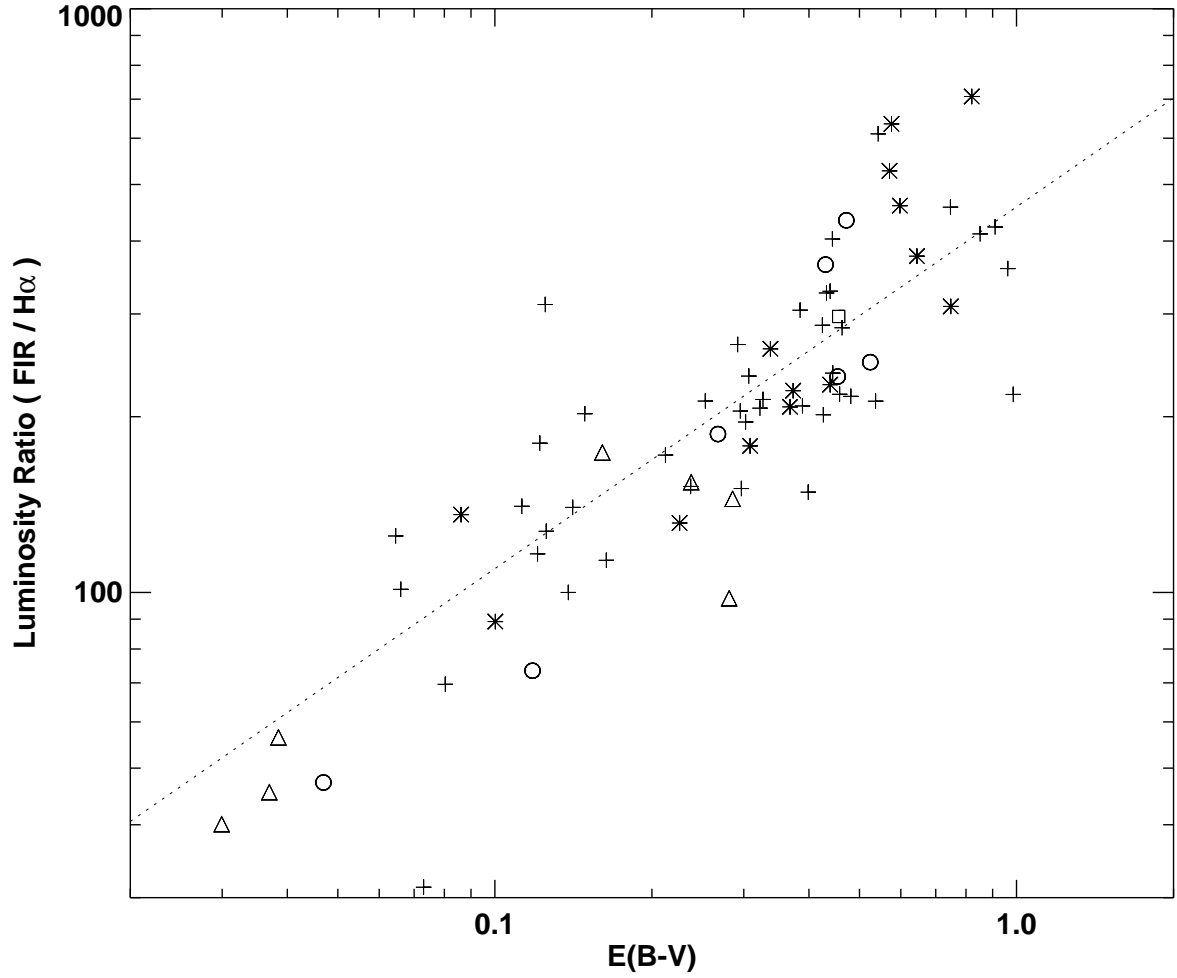}
\caption{The relationship
 between the reddening E($B-V$) derived from the Balmer
decrement F(H$\alpha$)/F(H$\beta$) and the ratio of the IR to \Ha\ 
luminosities.
\label{EB-V_vs_L_Ratio}}
\end{figure}

Figure~\ref{EB-V_vs_L_Ratio} shows the relationship between the
ratio of the two luminosities ${\rm L(FIR)/L(H\alpha)}$ and E($B-V$).    
The Spearman Rank correlation correlation coefficient is
0.86. The two-sided probability of obtaining a value of 0.86 by
chance is $\sim 1.2 \times 10^{-23}$, supporting the very strong 
correlation between ${\rm L(FIR)/L(H\alpha)}$ and E($B-V$).

\placefigure{EB-V_vs_SFR_Ratio}

The {\it fitexy} least squares line fitting routine 
gives 

\begin{equation}
\log\left[\frac{\rm L(FIR)}{\rm L(H\alpha)}\right] = (0.62\pm0.08)
\log [{\rm E}(B-V)] + 2.66 \pm 0.06.
\end{equation}

\begin{figure}
\plotone{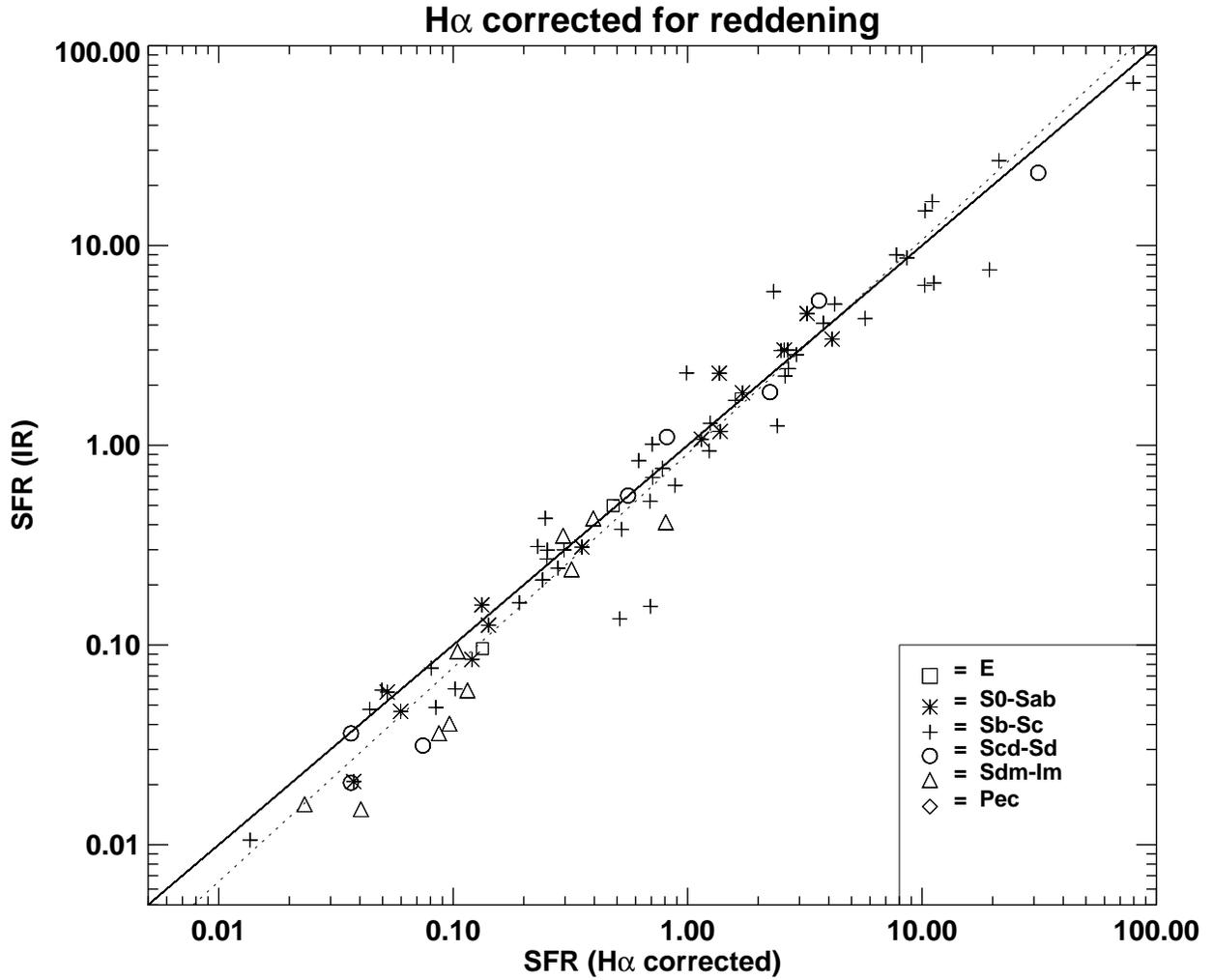}
\caption{As in Figure~1, with the \Ha\ flux  corrected
for reddening using the Whitford reddening curve as
parameterized by \citet{Miller72}.  The SFR(IR) and SFR(\Ha) are
nearly equal.  The legend indicates the Hubble types.
The galaxies with
the largest SFRs are late-type spirals. 
\label{SFR_Hacorr_vs_SFR_IR}}
\end{figure}

\noindent
This relationship is model independent.
If we use E($B-V$) to correct ${\rm L(H\alpha)}$ 
for extinction, the difference
between SFR(IR) and SFR(\Ha) could be reduced, supporting similar
results by \citet{Dopita02} and \citet{Rosa-Gonzalez02} for young
starforming galaxies.  Infact, after making the correction, we do 
arrive at much better agreement between the corrected \Ha\ star 
formation rate
(SFR(\Ha$_{\rm corr}$)) and the IR star formation rate
(Figure~\ref{SFR_Hacorr_vs_SFR_IR}).  This relation is striking
because many nearby field galaxies are cool IRAS sources.  Only 6 out
of the 81 galaxies in Figure~\ref{SFR_Hacorr_vs_SFR_IR} have a
60$\mu$m to 100$\mu$m flux ratio greater than 0.6, typical for objects
dominated by hot stars \citep{Bothun89}.

\placefigure{SFR_Hacorr_vs_SFR_IR}

We fit a straight line to the NFGS data in Figure~\ref{SFR_Hacorr_vs_SFR_IR}
using the same method as for Figure~\ref{SFR_Ha_vs_SFR_IR}.  The errors 
are $\sim30$\% for both SFR(\Ha) and SFR(IR).  
The resulting fit is

\begin{equation}
\log [{\rm SFR(IR)}] = (1.07\pm0.03) \log [{\rm SFR (H \alpha_{corr})}] -0.04\pm0.02
\end{equation}

or, 

\begin{equation}
{\rm SFR(IR)} = (0.91\pm0.04) {\rm SFR (H \alpha_{corr})^{1.07\pm0.03}}.
\end{equation}

The Spearman rank correlation coefficient for (SFR(\Ha$_{\rm corr}$))
and SFR(IR) is 0.98, with the 2-sided probability of obtaining 
this value of $\lesssim 10^{-30}$.  The rms scatter of the data around 
this line decreases to 0.12 in the log.
Clearly correcting \Ha\ for reddening results in a closer correspondence
between SFR(IR) and SFR(\Ha), and reduces the scatter observed
in Figure~\ref{SFR_Ha_vs_SFR_IR}.  

If the general stellar radiation field contributes significantly to
the IR emission in early-types, we would expect the late-types to
display a larger scatter than early types.  We find the contrary: the rms
scatter for the early-types (S0-Sab) is 0.08, 
compared to 0.13 for the late-types (Sb-Im).

\begin{figure}
\plotone{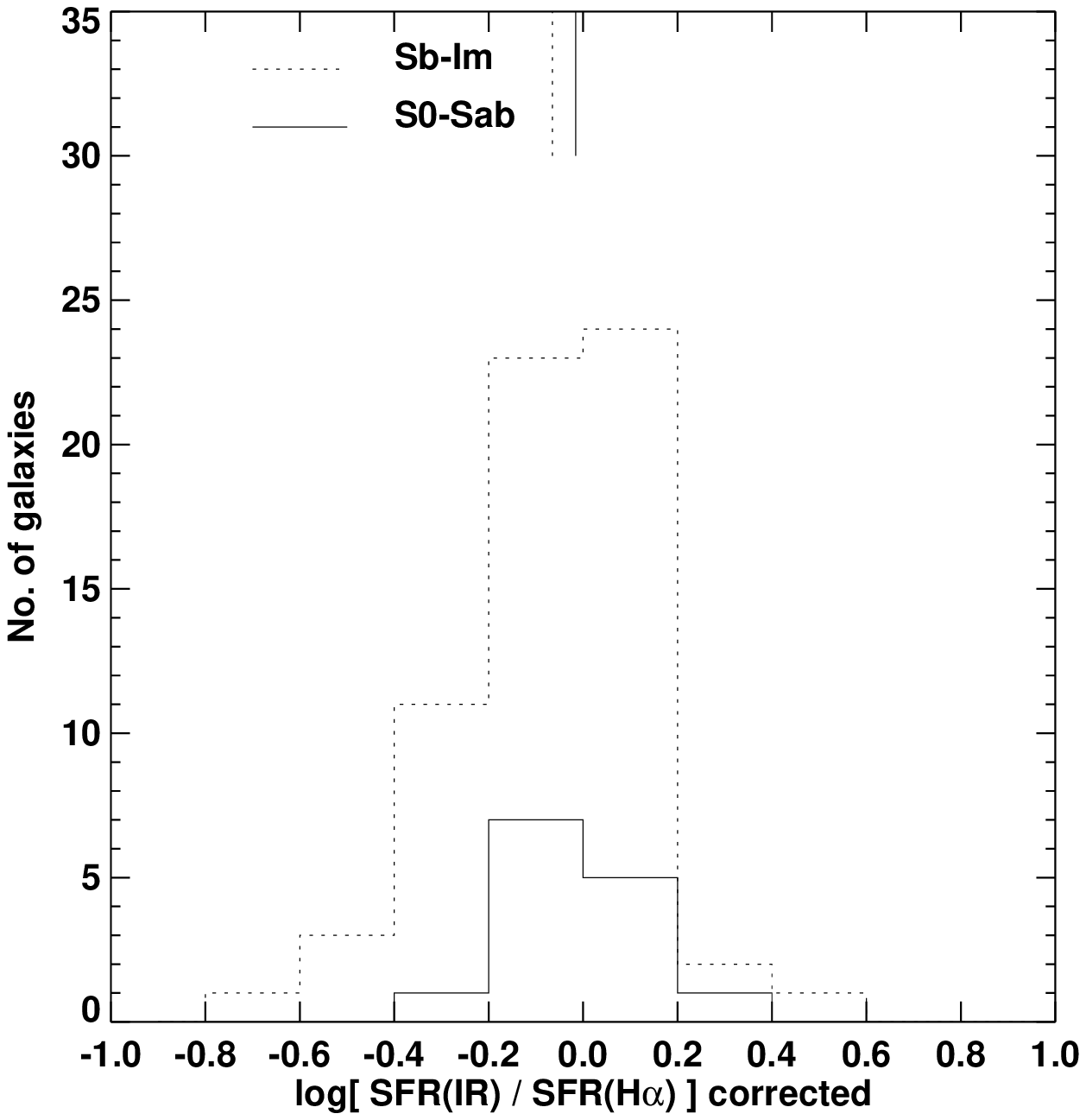}
\caption{The ratio of the infrared to \Ha\ star formation rates. 
The \Ha\ flux  has been corrected for
reddening.  The two vertical lines at the top of the Figure indicate
the mean of the early and late-type distributions.  
\label{hist_corr}}
\end{figure}

We plot a histogram of the logarithm of the ratio of the two SFRs, 
log(SFR(IR)/SFR(\Ha)), 
in Figure~\ref{hist_corr}.   After correction for reddening, the
means of the two SFR ratios for early and late types are approximately
equal: $-0.02\pm0.03$ and $-0.07\pm0.02$ respectively.  Greater reddening 
for the early types appears to explain the offset in 
Figure~\ref{hist_uncorr}.  The overall mean in the ratios is $-0.05\pm0.02$. 

\placefigure{hist_corr}

By selecting early-types with infrared detections, we bias
our early-type sample towards galaxies with more recent
star-formation than is normal in these galaxies.   The number of
early-types with 60 and 100 $\mu$m IRAS detections is 
17/59 (29\%), compared with the
number of late-types with detections 73/112 (65\%).  
This detection rate supports previous studies 
which showed that early-types in general have less recent star
formation than late types \citep[eg.,][]{Kennicutt83,Sauvage94,Bendo02}.  
The strong agreement and very small scatter between 
SFR(IR) and SFR(\Ha) 
for the infrared early-types implies that they are dominated in both \Ha\ 
and the IR by recent star formation.  

We obtain SFR estimates by multiplying the luminosity at the relevant
wavelength by a constant.  Thus, the SFR correlation 
(Figure~\ref{SFR_Hacorr_vs_SFR_IR}) is 
really equivalent to a correlation between the far-infrared 
and \Ha\ luminosities.  The line of best fit to the luminosity 
correlation is;

\begin{equation}
{\rm log\left[\frac{L(FIR)}{\rm L_{\odot}}\right]} =(1.07 \pm 0.02) {\rm \log\left[\frac{L(H\alpha_{corr})}{\rm L_{\odot}}\right]}] + (1.4 \pm 0.2)
\end{equation}

or,
\begin{equation}
\frac{\rm L(FIR)}{\rm L_{\odot}} = (25 \pm 12)\left[\frac{\rm L(H\alpha_{corr})}{\rm L_{\odot}}\right]^{(1.07 \pm 0.02)} \label{LIRLHa}
\end{equation}

For the K98 star formation rate indicators to produce equivalent 
IR and \Ha\ star formation rates,
the ratio ${\rm L(FIR)/L(H\alpha_{corr})}$ multiplied by the
ratio of the constants in the K98 relations (including the C00 factor of 1.75
to convert L(FIR) to L(IR))
should be $\sim 1$.  The mean ${\rm L(FIR)/L(H\alpha_{corr})}$ for
our sample is $96\pm4$.  Multiplying the mean ${\rm L(FIR)/L(H\alpha_{corr})}$ by ratio of the K98 relations gives 
$(96\pm4)(7.9\times10^{-44})/(7.9\times10^{-42}) = 0.96 \pm 0.04$.
Clearly, the combination of the C00 conversion and the ratio of
the K98 coefficients gives an excellent estimate of the relationship
between ${\rm L(FIR)/L(H\alpha_{corr})}$ and ${\rm SFR(IR)/SFR(H\alpha_{corr})}$. 

It is interesting that such a strong FIR-\Ha\ 
correlation occurs.
In the simplest picture, the correlation indicates 
that a single physical process is responsible for both the FIR and 
\Ha\ emission.  In reality, 
the FIR and \Ha\ radiation may be produced in quite different environments.
The \Ha\ radiation comes from gas ionized by the UV radiation 
from young, hot OB stars.  This process occurs within many \HII\ 
regions throughout a galaxy.   
The FIR radiation source is more complex.  

The far-IR spectra of galaxies result from large dust grains
of diameter 0.01-0.3.  The spectrum from these grains is characteristic
of thermal emission at 15-20K \citep[see][for a review]{Draine01},
and accounts for nearly all the emission at $\lambda > 60\mu$m.  
Some authors believe that an older stellar population 
can make a significant contribution to the far-IR emission
\citep[eg.,][]{Lonsdale87,Sauvage94,Mazzei94} for early-type
galaxies.  However, our \Ha -FIR luminosity correlation implies that the 
dominant UV and 
optical heating source for this emission is the young stellar population. 
This hypothesis is in agreement with various other physical correlations linking 
the FIR emission to young stellar populations, such as the radio-FIR
correlation \citep[eg.,][]{Gavazzi86}, and the 12$\mu$m-FIR luminosity
correlation \citep{Shapley01}.  Both of these correlations 
hold for all galaxy types.   

These correlations suggest that the gas and dust must be closely 
coupled in all FIR-bright galaxies.  \citet{Malhotra01} found a 
strong correlation between ${\rm L_{[OI]}/ L_{[CII]}}$
and F(60$\mu$m)/F(100$\mu$m), indicating that both gas and dust
temperatures increase together.   The most plausible explanation for
this effect is that the majority of the dust and gas heating in 
galaxies occurs very close to the actively star-forming regions. 

Modeling by \citet{Popescu02} for normal star-forming galaxies 
strengthens this explanation.
The Popescu et al. models predict that radiation from \HII\ regions is the dominant
energy source for the dust emitting at 60$\mu$m.  Even at 100$\mu$m,
\HII\ regions contribute $\sim1/3$ of the radiation source.  The other
$2/3$ is produced by the diffuse UV and optical radiation, which could
also contain some contribution from the young stellar population.  

Even if the FIR emission is dominated by the young stellar population,
one would expect the ratio of gas and dust masses to vary from one
\HII\ region to another, and from one galaxy to another.  
Similarly, the dust geometry and composition must vary.  
If these variations occur, they do not appear to have an 
effect on the relationship between the global FIR and \Ha\ luminosities.

\section{Conclusions \label{conclusions}}

We analyze a sample of 81 far-infrared detected galaxies from the Nearby Field 
Galaxy Survey.  The NFGS is an objectively selected sample of
galaxies in the CfA1 catalog.  The sample contains a representative
mix of all galaxy types and morphologies represented in the CfA1 survey.
In addition, the NFGS is unique in the quality and coverage of the
integrated spectra.
We have taken extensive care to ensure that the integrated \Ha\ flux for 
each galaxy is as reliable as possible.  We have corrected these fluxes
carefully for extinction and underlying stellar absorption.

To compare the IR and \Ha\ star formation rates, we make the following
standard assumptions: (1) the Whitford reddening law is appropriate 
for all galaxies in our sample, (2) the star formation rates are 
proportional to the relevant luminosities, and (3) that the IR SED of
the galaxies in our sample can be approximated by the \citet{Calzetti00}
relation.  

With the assumed reddening, the data provide a direct measurement of the
ratio of the IR and \Ha\ SFR constants which transform
the luminosities to star formation rates.  Correction
for reddening brings the IR and \Ha\ SFRs into 
agreement to within 10\% on average, and reduces the scatter.
 
Correction for reddening also brings the mean of SFR(IR)/SFR(H$\alpha$)
for early and late types into agreement.  One average, early types in this
sample are more heavily reddened than late types.
It is intriguing that the ratio between the SFR(IR) and SFR(\Ha)
is the same for early and late types.  The IR emission from 
early-type spirals
may contain a significant contribution from the general stellar
radiation field.  This contribution should increase the observed IR 
luminosity. \citet{Inoue02} suggested that this effect is offset by a 
lower dust opacity.  This compensation would require a remarkable conspiracy
to hold for all galaxy types.
The physical properties of dust and the nature of the general stellar
radiation field probably vary from galaxy to galaxy.  The
variation in the general stellar radiation field should result
in a larger dispersion around
the SFR relation for early types than late-types.  
We do not observe this variation.  Rather, the infrared 
NFGS early-type galaxies have a smaller dispersion than the
late-types.

Previous studies have shown that early-type galaxies contain a lower
fraction of young-to-old stellar populations than late-types
\citep[eg.,][]{Kennicutt83,Sauvage94,Davis83,Huchra83}.  
This result does not determine what combination of
young-to-old populations produces the IR emission in each galaxy.
The agreement of SFR(IR) with SFR(\Ha) (or L(FIR) with L(\Ha)) 
for both early types and
late-types implies that the FIR emission results from the same young
stellar population which produces the \Ha\ emission.
This hypothesis is supported by other physical correlations linking
the FIR emission to the young stellar population, including
the radio-FIR correlation\citep{Gavazzi86} and the 12$\mu$m-FIR luminosity
correlation \citep{Shapley01}.
These correlations suggest that the dust is heated in close
proximity to the young star-forming regions.  In addition, the
FIR-\Ha\ luminosity correlation implies that the relationship between the global dust 
and gas properties is universal for all FIR-detected galaxies in the NFGS. 

Further observations and modeling are required to prove that the young 
stellar population is responsible for FIR emission in all galaxy types, 
rather than a compensation effect as suggested by \citet{Inoue02}.
Near-infrared observations would enable separation of the young and old
stellar populations. The Space Infrared Telescope Facility (SIRTF) will 
help determine whether the global dust and gas properties are truly 
universal for all FIR-detected galaxies in the NFGS.

\acknowledgments
We enjoyed discussions about many aspects of this paper with 
Rob Kennicutt, Dan Fabricant, and Stephane Charlot.
L. J. Kewley is supported by a Harvard-Smithsonian 
CfA Fellowship.   M. J. Geller is supported by the Smithsonian Institution.

\clearpage

\clearpage 

\clearpage

\clearpage 

\begin{deluxetable}{llrlrrrrrrrrrr}
\tabletypesize{\scriptsize}
\tablecaption{The NFGS Sample with 60 and 100$\mu$m IRAS detections 
\label{sample_table}}
\tablehead{{ID}
& Name
& cz 
& Type
& E(B-V)
& F$_{60}$\tablenotemark{a}
& F$_{100}$\tablenotemark{a}
& $\log \left[\frac{\rm L_{corr}(H\alpha)}{\rm L_{\odot}}\right]$
& $\log \left[\frac{\rm L(FIR)}{\rm L_{\odot}} \right]$
& SFR
& SFR
& SFR\\
& & km/s& & & Jy & Jy & & &\Ha & \Ha$_{\rm corr}$ & IR}
\startdata
4   & UGC 00439 & 5302 & Sa   & 0.44 & 1.14  & 2.80 &   8.13 & 10.04 &  1.50 &  4.13 &  3.40 \\
5   & UGC 00484 & 4859 & Sb   & 0.37 & 0.91 & 2.50 &   7.94 & 9.90  &  1.15 &  2.69 &  2.42 \\
16  & UGC 01154 & 7756 & Sbc  & 0.43 & 0.56 & 1.96 &   8.27 & 10.15 &  2.14 &  5.72 &  4.30 \\
17  & UGC 01155 & 3158 & Sbc  & 0.30 & 0.49 & 1.21 &   7.35 &  9.23 &  0.35 &  0.69 &  0.52 \\
19  & NGC 695   & 9705 & Sc   & 0.75 & 7.69  & 12.84 &   9.41 & 11.33 & 14.27 & 79.67 & 65.02  \\
21  & UGC 01551 & 2669 & Sdm  & 0.28 & 0.48  & 1.50 & 7.42 &  9.13 &  0.42 &  0.81 &  0.41  \\
23  & IC 197    & 6332 & Sbc  & 0.42 & 0.97  & 2.31 & 8.09 & 10.12 &  1.43 &  3.79 &  4.08  \\
25  & UGC 01630 & 4405 & Sb   & 0.44 & 1.55  & 3.26 & 7.91 &  9.99 &  0.91 &  2.50 &  2.98  \\
27  & NGC 927   & 8258 & Sc   & 0.54 & 0.82 & 2.29 & 8.52 & 10.31 &  2.98 & 10.26 &  6.32  \\
28  & UGC 01945 & 1762 & Sdm  & 0.24 & 0.24 & 0.79 & 6.53 &  8.48 &  0.06 &  0.10 &  0.09  \\
37  & UGC 04713 & 9036 & Sb   & 0.96 & 0.61 & 2.81 & 8.80 & 10.39 &  2.11 & 19.37 &  7.56  \\
39  & NGC 2780  & 1951 & Sab  & 0.34 & 0.36 & 1.04 & 6.63 &  8.71 &  0.06 &  0.13 &  0.16  \\
43  & NGC 2844  & 1486 & Sa   & 0.37 & 0.48 & 1.46 & 6.66 &  8.61 &  0.06 &  0.14 &  0.13  \\
44  & NGC 3011  & 1517 & S0/a & 0.09 & 0.23 & 0.61 & 6.23 &  8.28 &  0.04 &  0.05 &  0.06  \\
45  & NGC 3009  & 4666 & Sc   & 0.30 & 0.27 & 0.82 & 7.36 &  9.35 &  0.35 &  0.71 &  0.69  \\
46  & IC 2520   & 1226 &../Pec& 0.46 & 3.48  & 6.73 & 7.19 &  9.21 &  0.17 &  0.48 &  0.50  \\
47  & UGC 05354 & 1172 & Sm   & 0.04 & 0.43 & 0.92 & 6.57 &  8.29 &  0.11 &  0.11 &  0.06  \\
48  & NGC 3075  & 3566 & Sc   & 0.15 & 0.79 & 1.72 & 7.36 &  9.52 &  0.50 &  0.71 &  1.01  \\
49  & UGC 05378 & 4185 & Sb   & 0.46 & 0.48 & 1.30 & 7.60 &  9.48 &  0.43 &  1.24 &  0.94  \\
50  & NGC 3104  &  604 & Im   & 0.03 & 0.37 & 1.00 & 6.12 &  7.69 &  0.04 &  0.04 &  0.02  \\
53  & UGC 05522 & 1228 & Sc   & 0.08 & 0.26 & 0.85 & 6.44 &  8.20 &  0.07 &  0.08 &  0.05  \\
56  & NGC 3213  & 1412 & Sbc  & 0.40 & 0.25 & 0.79 & 6.52 &  8.29 &  0.04 &  0.10 &  0.06  \\
57  & NGC 3264  &  929 & Sdm  & 0.00 & 0.46 & 1.02 & 6.49 &  8.12 &  0.11 &  0.11 &  0.04 \\
58  & NGC 3279  & 1422 & Sc   & 0.54 & 1.83  & 5.37 & 6.90 &  9.15 &  0.07 &  0.25 &  0.43  \\
59  & UGC 05744 & 3338 & Sc   & 0.29 & 0.86 & 1.48 & 7.30 &  9.44 &  0.32 &  0.62 &  0.84  \\
60  & UGC 05760 & 2997 & Scd  & 0.43 & 1.05  & 3.09 & 7.42 &  9.55 &  0.30 &  0.82 &  1.10  \\
61  & IC 2591   & 6755 & Sbc  & 0.20 & 0.59 & 1.58 & 7.94 &  9.99 &  1.68 &  2.67 &  2.99 \\
63  & UGC 05798 & 1534 & Sc   & 0.13 & 0.36 & 0.62 & 6.42 &  8.40 &  0.06 &  0.08 &  0.08  \\
65  & CGCG 212-039 & 10693 & Sbc  & 0.64 & 0.43 & 1.32 & 7.88 & 10.28 &  0.53 &  2.33 &  5.88  \\
73  & NGC 3510  &  704 & Sd   & 0.05 & 0.60 & 1.46 & 6.38 &  8.01 &  0.07 &  0.07 &  0.03  \\
76  & IRAS F11041+5127 & 2204 & Sc   & 0.06 & 0.44 & 1.49 & 6.91 &  8.94 &  0.22 &  0.25 &  0.27  \\
77  & IC 673    & 3851 & Sa   & 0.37 & 0.59 & 1.90 & 7.57 &  9.54 &  0.49 &  1.15 &  1.07  \\
82  & NGC 3633  & 2553 & Sa   & 0.82 & 3.08  & 5.34 & 7.75 &  9.78 &  0.26 &  1.72 &  1.83  \\
88  & UGC 06575 & 1225 & Sc   & 0.11 & 0.29 & 0.76 & 6.16 &  8.19 &  0.03 &  0.04 &  0.05  \\
89  & NGC 3795  & 1154 & Scd  & 0.00 & 0.23 & 0.68 & 6.08 &  8.07 &  0.05 &  0.05 &  0.04 \\
90  & UGC 06625 &10964 & Sc   & 0.28 & 1.23   & 3.39 & 8.56 & 10.73 &  5.82 & 11.06 & 16.59 \\
91  & NGC 3795A & 1091 & Sc   & 0.27 & 0.40   & 1.34 & 6.21 &  8.29 &  0.03 &  0.05 &  0.06 \\
96  & UGC 06805 & 1033 & S0   & 0.23 & 0.52   & 0.73  & 6.29 &  8.18 &  0.04 &  0.06 & 0.05  \\
98  & IC 746    & 5027 & Sb   & 0.30 & 0.50   & 1.13  & 7.61 &  9.62 &  0.63 &  1.25 & 1.29  \\
100 & NGC 3978  & 9978 & Sbc  & 0.38 & 2.24  & 6.88 & 8.84 & 10.94 &  8.77 & 21.27 & 26.61  \\
105 & UGC 07020A& 1447 & S0   & 0.31 & 1.82  & 2.29 & 7.06 &  9.00 &  0.17 &  0.35 & 0.31  \\
107 & NGC 4120  & 2251 & Sc   & 0.14 & 0.47  & 1.58 & 6.98 &  8.99 &  0.21 &  0.30 & 0.30  \\
109 & NGC 4141  & 1980 & Sc   & 0.07 & 0.31 & 0.85 & 7.22 &  8.64 &  0.43 &  0.51 & 0.14  \\
110 & NGC 4159  & 1761 & Sdm  & 0.29 & 0.81 & 1.56 & 7.02 &  8.89 &  0.17 &  0.32 & 0.24  \\
112 & NGC 4238  & 2771 & Sc   & 0.14 & 0.49 & 1.08 & 7.23 &  9.09 &  0.38 &  0.52 & 0.38  \\
114 & UGC 07358 & 3639 & Sc   & 0.32 & 0.46 & 1.55 & 7.40 &  9.40 &  0.37 &  0.78 & 0.77  \\
122 & UGC 07690 &  540 & Sdm  & 0.01 & 0.54  & 1.21 & 5.88 &  7.72 &  0.02 &  0.02 & 0.02  \\
124 & NGC 4509  &  907 & Sm   & 0.04 & 0.61  & 0.52  & 6.45 &  8.07 &  0.08 &  0.09 & 0.04  \\
125 & UGC 07761 & 6959 & Sb   & 0.43 & 0.96 & 2.50 & 8.14 & 10.22 &  1.57 &  4.24 & 5.09  \\
127 & NGC 4758  & 1244 & Sbc  & 0.39 & 0.95 & 2.53 & 6.79 &  8.72 &  0.08 &  0.19 & 0.16  \\
133 & UGC 08231 & 2460 & Sb   & 0.07 & 0.36 & 0.97 & 6.96 &  8.90 &  0.24 &  0.28 & 0.24  \\
134 & IC 4213   &  815 & Scd  & 0.12 & 0.29 & 0.71 & 6.08 &  7.82 &  0.03 &  0.04 & 0.02  \\
135 & UGC 08400 & 3396 & Scd  & 0.27 & 0.42 & 1.22 & 7.26 &  9.26 &  0.30 &  0.56 & 0.56  \\
140 & NGC 5230  & 6855 & Sc   & 0.33 & 1.56   & 4.71 & 8.45 & 10.45 &  4.06 &  8.61 & 8.67  \\
141 & UGC 08630 & 2364 & Sm   & 0.16 & 0.60   & 1.44  & 6.98 &  9.06 &  0.20 &  0.29 & 0.35  \\
142 & NGC 5267  & 5941 & Sb   & 0.91 & 0.22   & 1.12 & 7.89 &  9.61 &  0.30 &  2.41 & 1.25  \\
144 & NGC 5338  &  777 & S0   & 0.75 & 0.39 & 0.63 & 6.09 &  7.83 &  0.01 &  0.04 & 0.02  \\
145 & NGC 5356  & 1397 & Sb   & 0.99 & 0.54 & 2.40 & 7.35 &  8.71 &  0.07 &  0.69 & 0.16  \\
148 & NGC 5425  & 2062 & Sc   & 0.24 & 0.37 & 1.40 & 6.89 &  8.84 &  0.14 &  0.24 & 0.21  \\
151 & NGC 5491  & 5845 & Sc   & 0.46 & 0.68 & 2.19 & 7.98 &  9.97 &  1.00 &  2.91 & 2.84  \\
153 & NGC 5541  & 7698 & Sc   & 0.44 & 2.63  & 5.12 & 8.52 & 10.69 &  3.71 & 10.30 & 14.91  \\
156 & UGC 09356 & 2234 & Sc   & 0.12 & 0.40 & 0.97 & 6.89 &  8.84 &  0.18 &  0.24 & 0.21  \\
158 & NGC 5762  & 1788 & Sa   & 0.10 & 0.20 & 0.74 & 6.59 &  8.44 &  0.10 &  0.12 & 0.08  \\
159 & UGC 09560 & 1215 & Pec  & 0.00 & 0.71 & 1.24 & 6.64 &  8.49 &  0.15 &  0.15 & 0.10 \\
162 & UGC 09660 &  626 & Sc   & 0.16 & 0.26 & 0.61 & 5.64 &  7.54 &  0.01 &  0.01 & 0.01  \\
164 & IC 1100   & 6561 & Scd  & 0.47 & 1.03  & 3.18 & 8.07 & 10.24 &  1.23 &  3.63 & 5.30  \\
165 & NGC 5874  & 3128 & Sbc  & 0.48 & 0.47 & 1.85 & 7.46 &  9.31 &  0.29 &  0.88 & 0.63  \\
166 & NGC 5875A & 2470 & Sc   & 0.12 & 0.51 & 1.11 & 6.87 &  9.01 &  0.17 &  0.23 & 0.31  \\
168 & IC 1124   & 5242 & Sab  & 0.60 & 0.95 & 2.71 & 7.92 &  9.99 &  0.65 &  2.58 & 2.99  \\
170 & UGC 09896 & 6461 & Sc   & 0.21 & 0.28 & 1.17  & 7.71 &  9.74 &  0.98 &  1.60 & 1.68  \\
172 & IC 1141   & 4458 & S0/a & 0.58 & 1.34 & 2.00 & 7.65 &  9.87 &  0.36 &  1.36 & 2.29  \\
174 & NGC 6007  &10548 & Sbc  & 0.31 & 0.70 & 2.03  & 8.40 & 10.47 &  3.84 &  7.78 & 8.98  \\
175 & UGC 10086 & 2191 & Sc   & 0.25 & 0.64 & 1.28 & 6.91 &  8.99 &  0.14 &  0.25 & 0.30  \\
179 & NGC 6131  & 5054 & Sc   & 0.12 & 0.72 & 2.42  & 7.51 &  9.87 &  0.74 &  0.99 & 2.30  \\
184 & NGC 7328  & 2827 & Sab  & 0.64 & 1.16  & 3.95 & 7.65 &  9.58 &  0.31 &  1.38 & 1.17  \\
186 & UGC 12178 & 1925 & Sdm  & 0.15 & 1.08  & 2.72 & 7.11 &  9.15 &  0.28 &  0.40 & 0.43 \\
187 & UGC 12265 & 5682 & S0   & 0.57 & 1.62  & 2.53 & 8.02 & 10.17 &  0.87 &  3.23 & 4.57  \\
189 & NGC 7460  & 3296 & Sb   & 0.44 & 1.82  & 4.98 & 7.93 &  9.86 &  0.94 &  2.61 & 2.22  \\
192 & UGC 12519 & 4380 & Sd   & 0.45 & 0.76 & 2.59 & 7.86 &  9.78 &  0.79 &  2.25 & 1.85  \\
193 & NGC 7620  & 9565 & Scd  & 0.52 & 2.41  & 5.74  & 9.01 & 10.88 &  9.35 & 31.29 & 23.13  \\
195 & IC 1504   & 6306 & Sb   & 0.85 & 1.40  & 4.12 & 8.56 & 10.32 &  1.58 & 11.23 & 6.48  \\
\tablenotetext{a}{IRAS Fluxes at 60 and 100$\mu$m in units of Janskies}
\enddata
\end{deluxetable}

\end{document}